\documentclass[journal=nalefd,manuscript=letter,layout=twocolumn]{achemso}

\usepackage[version=3]{mhchem} 
\usepackage{float}
\restylefloat{table}
\usepackage{pdfpages}

\setkeys{acs}{maxauthors = 0}      

\setkeys{acs}{articletitle = true}

\author{Bassim~Arkook}
\affiliation{Physics and Astronomy, University of California, Riverside, CA 92521}
\author{Christopher~Safranski}
\affiliation{Physics and Astronomy, University of California, Irvine, CA 92697}
\author{Rodolfo~Rodriguez}
\affiliation{Physics and Astronomy, University of California, Riverside, CA 92521}
\author{Ilya~N.~Krivorotov}
\affiliation{Physics and Astronomy, University of California, Irvine, CA 92697}
\author{Tobias~Schneider}
\affiliation{Institute of Ion Beam Physics and Materials Research,
Helmholtz-Zentrum Dresden~-~Rossendorf e.V., 01328 Dresden, Germany}
\author{Kilian Lenz}
\affiliation{Institute of Ion Beam Physics and Materials Research,
Helmholtz-Zentrum Dresden~-~Rossendorf e.V., 01328 Dresden, Germany}
\author{J\"{u}rgen~Lindner}
\affiliation{Institute of Ion Beam Physics and Materials Research,
Helmholtz-Zentrum Dresden~-~Rossendorf e.V., 01328 Dresden, Germany}
\author{Houchen Chang}
\affiliation{Department of Physics, Colorado State University, Fort Collins, CO 80523}
\author{Mingzhong Wu}
\affiliation{Department of Physics, Colorado State University, Fort Collins, CO 80523}
\author{Yaroslav~Tserkovnyak}
\affiliation{Physics and Astronomy, University of California, Los Angeles, CA 90095}
\author{Igor~Barsukov}
\affiliation{Physics and Astronomy, University of California, Riverside, CA 92521}
\email{igorb@ucr.edu}

\title{Thermally driven two-magnet nano-oscillator with large spin-charge conversion}


\begin{document}

\begin{tocentry}
\begin{center}
\includegraphics[width=0.80\textwidth]{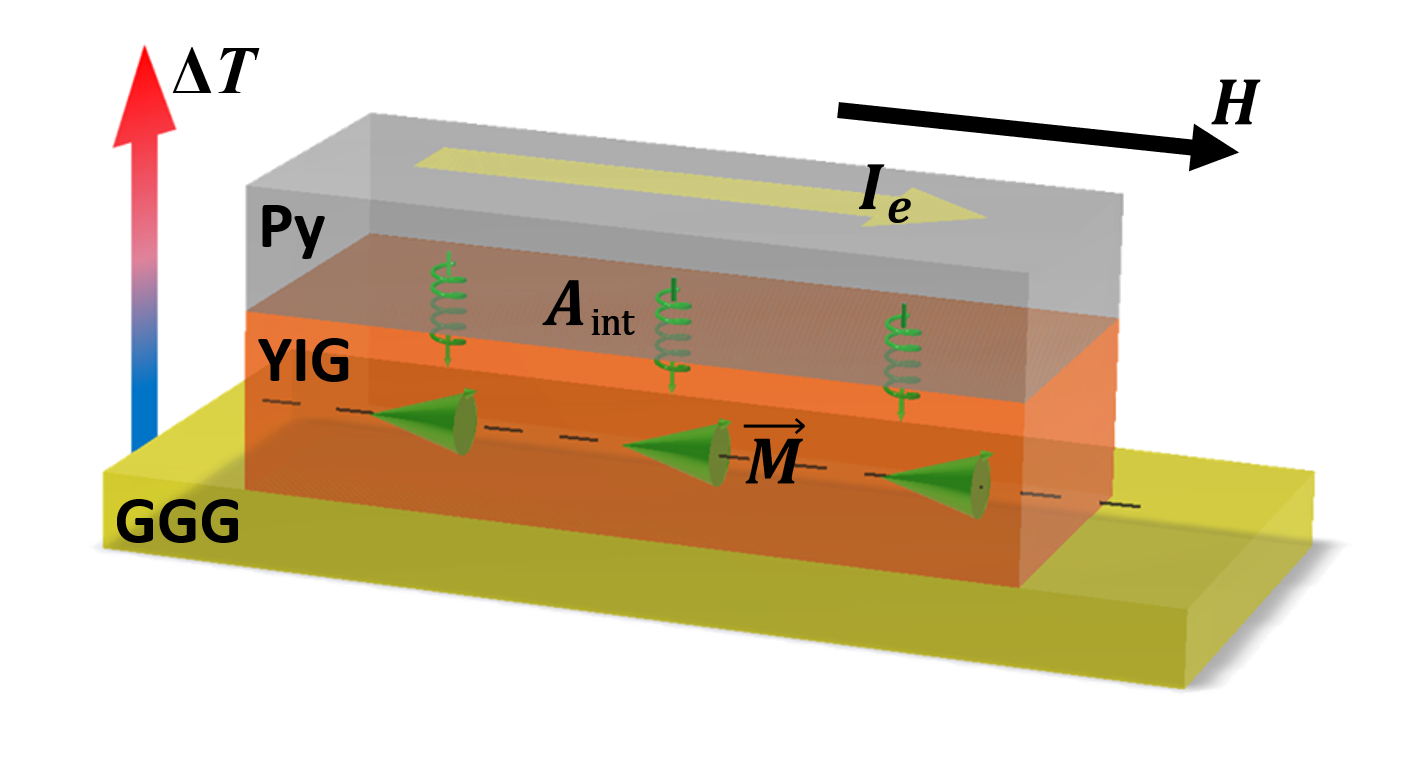}
\end{center}
\end{tocentry}

\begin{abstract}
Next-generation spintronic applications require material properties that can be hardly met by one material candidate. Here we demonstrate that by combining insulating and metallic magnets, enhanced spin-charge conversion and energy-efficient thermal spin currents can be realized. We develop a nanowire device consisting of an yttrium iron garnet and permalloy bi-layer. An interfacial temperature gradient drives the nanowire magnetization into auto-oscillations at gigahertz frequencies. Interfacial spin coupling and magnetoresistance of the permalloy layer translate spin dynamics into sizable microwave signals. The results show prospect for energy-efficient spintronic devices and present an experimental realization of magnon condensation in a heterogeneous magnetic system.  
\end{abstract}

\section{~}
Energy-efficient control of spin dynamics\cite{kiselev2003microwave} and sizable spin-charge conversion\cite{HellmanInterfaces} are central topics of spintronic research. Magnetic insulators have recently risen as promising material candidates for spintronic applications\cite{avci2017current, bozhko2016supercurrent, yang2018fmr}. They possess low magnetic damping and thus reduce energy dissipation\cite{wunderlich2017spintronics, collet2016generation}. The low thermal conductivity of magnetic insulators, once viewed as a challenge for heat management, bears an opportunity for spintronic devices via heat recycling. Interfacial heat flow in ultra-thin bi-layers of the ferrimagnetic insulator, yttrium iron garnet (YIG), and platinum has recently been shown to inject a spin current sufficient to induce auto-oscillations of YIG magnetization\cite{safranski2017spin}, creating a spin-torque oscillator driven by waste heat. 

While magnetic insulators help to avoid shunting of electrical currents, they possess no intrinsic means of spin-charge conversion and rely on proximity-mediated effects\cite{HellmanInterfaces,lee2016magnetic}. In YIG/Pt systems, the spin Hall magnetoresistance allows for read-out of spin information. However, it is rather small (typically 10$^{-1}$-10$^{-2}$\,\%)\cite{nakayama2013spin,safranski2017spin}, which presents an obstacle for integration of insulators in reliable spintronic applications.  

\begin{figure*}[ht]
\includegraphics[width=0.99\textwidth]{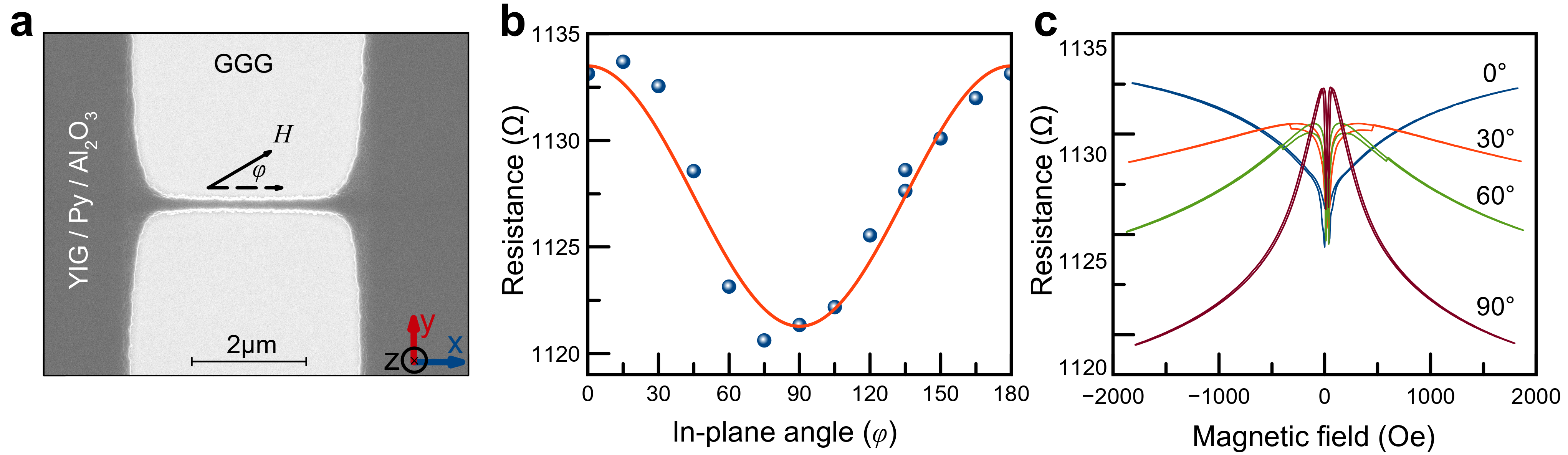}
\caption{Magnetoresistance of the nanowire device. (a)~Scanning electron micrograph of the nanowire. $H$ is the external magnetic field, applied in the film plane at angle $\varphi$ with respect to the nanowire axis. (b)~Device resistance as a function of field angle at $H=1.6$\,kOe. (c)~Device resistance as a function of the magnetic field for different in-plane angles $\varphi$. All measurements in this work are carried out at 77\,K.}
\label{fig1}
\end{figure*}

Metallic ferromagnets, on the other hand, possess intrinsic, sizable spin-charge effects based on spin-orbit interaction, which can be employed for electrical read-out of static and dynamic magnetic states\cite{SinovaSHE, tulapurkar2005spin, baumgaertl2016magnetization}. Recently, these spin-charge effects have received much attention as tunable sources of spin currents\cite{taniguchi2015spin, davidson2019perspectives, AminInterface}. Spin injection using anomalous Hall effect\cite{RalphAHE, DasAHE, TulapurkarAHE}, planar Hall effect\cite{safranski2019spin}, rotational-symmetry spin-orbit effect\cite{HumphriesRotSym} in multilayers, and spin-orbit torques in a single ferromagnetic layer\cite{haidar2019single} have been experimentally realized.

Here, we propose to exploit the virtues of insulating and metallic magnets and fabricate nanowires from thin film bi-layers of YIG and permalloy (Py=Ni$_{80}$Fe$_{20}$). We find that spin coupling at the metal-insulator interface and anisotropic magnetoresistance (AMR) in Py allow for electrical read-out of spin dynamics in YIG. Furthermore, application of interfacial temperature gradient from ohmic heating drives the nanowire into magnetic auto-oscillations which translate into sizable microwave signals. Our findings demonstrate that nanoscale heterostructures based on a combination of metallic and insulating magnets offer both sizable spin-charge conversion and thermal spin-torques, thus enabling next-generation energy-efficient spintronic applications. The observed auto-oscillations, moreover, show an experimental realization of magnon condensation \cite{demokritov2006bose} in a two-magnet system with hybridized spin wave modes.

Using sputtering deposition\cite{liu2014ferromagnetic}, we prepare multilayer thin films consisting of \ce{Gd3Ga5O12}(GGG substrate)/YIG(20\,nm)/Py (5\,nm)/AlOx(2\,nm). The top layer prevents Py from oxidation. By means of negative e-beam lithography and ion milling, we fabricate nanowire devices (180\,nm width, 3.4\,$\mu$m  length) which fan out into sub-millimeter\cite{safranski2017spin} large electric leads (Fig.\,1a).

We apply a large magnetic field $H=1.6$\,kOe in the sample plane and measure the device resistance as a function of field angle $\varphi$ with respect to the nanowire axis (Fig.\,1b). All measurements are carried out in a cryostat at 77\,K thermal bath temperature. As expected for the anisotropic magnetoresistance of Py\cite{duan2014nanowire}, the resistance follows $R=R_0+\Delta R \cos(2\varphi)/2$ behavior, where $R_0=1127\,\Omega$ and $\Delta R / R_0 = 1.24\%$ is the magnetoresistance ratio. Figure\,1c shows the device resistance as a function of the magnetic field. Again as expected\cite{duan2014nanowire}, in the hard axis (at $\varphi=90$\,deg due to the shape anisotropy of the nanowire) the curve is bell-shaped and in the easy axis ($\varphi=0$\,deg) the curve is V-shaped. However, large resistance drops near zero-field and incomplete saturation in the easy axis suggest some coupling of YIG and Py spins at the interface.

\begin{figure*}[ht]
\includegraphics[width=0.78\textwidth]{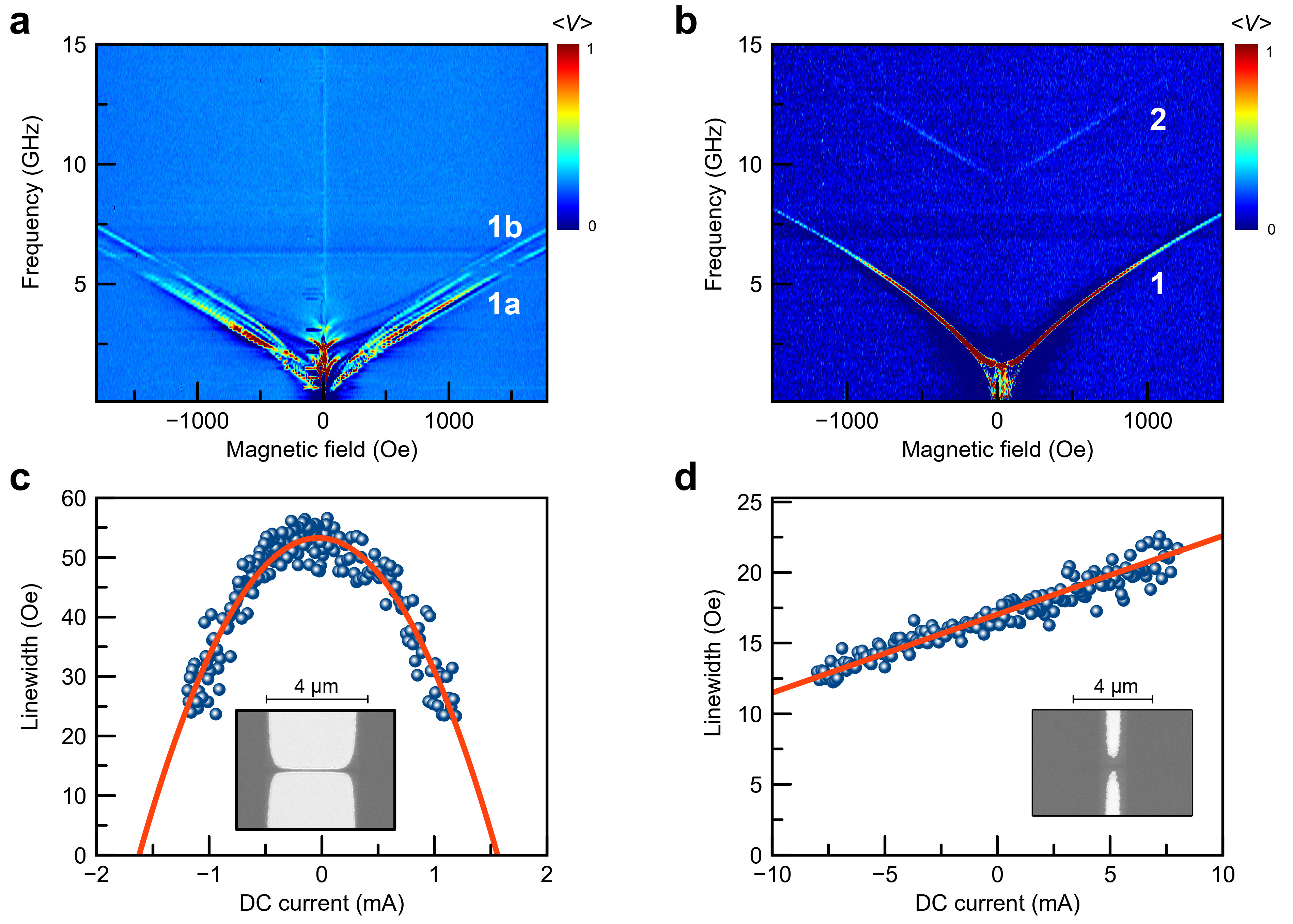}
\caption{Spin-torque ferromagnetic resonance. (a)~Spin wave spectrum of the long nanowire near the hard axis. The normalized ST-FMR signal $\langle V\rangle$ is color-coded. (b)~Spin wave spectrum of the short nanowire near the hard axis. (c)~Linewidth of the lowest spin wave mode for the long nanowire in the easy axis at $f=6.8$\,GHz. (d)~Linewidth of the lowest spin wave mode of the short nanowire $\sim$20\,deg off easy axis at $f=2.5$\,GHz. Red curves are guides to the eye.}
\label{fig2}
\end{figure*}

The presence of interfacial spin coupling is anticipated to have an impact on the spin wave spectrum of the device\cite{sluka2019emission,PhysRevApplied.11.034065,liu2018long,PhysRevB.34.7788,NOGUES1999203,PhysRevLett.120.127201,PhysRevLett.120.217202}. While zero coupling should leave individual spin wave spectra of YIG and Py layers nearly unchanged (except for the dipolar interaction), strong coupling should significantly hybridize and delocalize the spin waves over both, YIG and Py, layers and can lead to spin wave modes of acoustic and optical type\cite{sluka2019emission}.

To evaluate the spin wave spectrum, we carry out spin-torque ferromagnetic resonance (ST-FMR) measurements near the hard axis with field modulation\cite{gonccalves2013spin} on the nanowire shown in Fig.\,1a. Figure~2a presents one group of spin waves as a function of frequency and field. This group is dominated by two modes labeled '1a' and '1b'. The spin wave spectrum is not only determined by the coupling between YIG and Py, but also by the nanowire dimensions. We thus carry out ST-FMR on a much shorter nanowire (displayed in inset of Fig.\,2d). As shown in Fig.\,2b, the frequency-field data presents one spin wave mode, labeled '1', similar to the spin waves observed in the longer nanowire. At higher frequencies, another spin wave mode, labeled '2' is found.

To assess the nature of these spin waves, we perform micromagnetic simulations using typical material parameters of YIG and Py (Supporting Information). We find the best agreement between experiment and micromagnetics for $A_{\mathrm{int}}=0.4$\,pJ/m -- a moderately large, ferromagnetic-type coupling at the interface. The spin waves of the lower-frequency group ('1a' and '1b') are predominantly localized in the YIG layer. They correspond to the normal spin wave eigenmodes\cite{PhysRevB.92.104424,duan2014nanowire} of a nanowire (see Supporting Information) -- the number of nodes increases with increasing frequency. The higher-frequency mode ('2') corresponds to the lowest-order eigenmode of the nanowire, but is predominantly localized in the Py layer. Due to the symmetry of excitation field in the FMR experiment (Oersted field of the Py layer), the group '2' and higher-order spin waves of the group '1' are very weakly excited\cite{duan2014nanowire}, and thus not visible in Fig.\,2a.  

Micromagnetic analysis reveals that spin wave modes at lower frequencies, while delocalized over both layers, are more strongly excited in one layer (see Supporting Information). For instance, the mode '1a' is predominantly localized in YIG, but also drags the magnetization of Py. Such delocalization allows for detecting this mode electrically via magnetoresistance of Py. While at low frequencies, the magnetizations of YIG, $\vec{M}_{\textrm{YIG}}$, and Py, $\vec{M}_{\textrm{Py}}$, precess nearly in-phase (acoustic modes), with increasing frequency the phase difference approaches $\pi$ (optical modes). Moreover, the delocalization increases, and the spin wave mode become strongly hybridized.

To explore the possibility of manipulating the spin dynamics of the two-magnet devices, we analyze the ST-FMR linewidth, which is representative of the effective damping. The linewidth of the lowest mode '1a' of the long nanowire is shown in Fig.\,2c as a function of electric DC current sent through the Py layer. The linewidth decreases for both polarities of the bias current in the easy axis  $\varphi=0\,$deg. The observed decrease of the damping cannot be explained by any known spin-orbit torque generated by electric current in Py\cite{davidson2019perspectives} (Supporting Information): (i)~In particular, the spin Hall effect produces a spin current with polarization $\vec{\sigma}$ parallel to $\hat y$ and thus\cite{SinovaSHE} does not contribute to damping-like torque ($\propto\vec{M}_{\textrm{YIG}}\times\vec{M}_{\textrm{YIG}}\times\vec{\sigma}$) in easy axis. (ii)~The Anomalous Hall torque\cite{RalphAHE, DasAHE, TulapurkarAHE} is zero for parallel orientation of electric current and Py magnetization, which is the case for easy axis. (iii)~The Planar Hall torque\cite{safranski2019spin} is zero when magnetization lies in the film plane. (iv)~The Rotational-Symmetry torque proposed in Refs.\cite{HumphriesRotSym,davidson2019perspectives} is zero for parallel orientation of YIG and Py magnetizations. Furthermore, spin-orbit torques are generally odd in electric current and cannot explain the symmetric behavior shown in Fig.\,2c.

On the other hand, ohmic heating in the Py layer is nearly quadratic in electric current. Using finite-element simulations (Supporting Information), we estimate a notable temperature gradient across the layers of $\sim0.3\,$K/nm at $\sim1.8$\,mA for the long nanowire. The temperature gradient generates an anti-damping spin-torque via spin Seebeck effect\cite{safranski2017spin,ohnuma2015magnon,bender2014dynamic,WuSpinSeebeckPRL}, that is independent of the DC current polarity, consistent with the data in Fig.\,2c. In the short nanowire, the heat easily dissipates into the adjacent electrical leads, which reduces the temperature gradient to $\sim0.004\,$K/nm. With such a small temperature gradient, the spin Seebeck torque is negligible. The linewidth of the lowest spin wave mode in Fig.\,2c therefore depends linearly on the electric current, which is characteristic for spin-orbit torques (described above).

\begin{figure*}[h]
\includegraphics[width=0.80\textwidth]{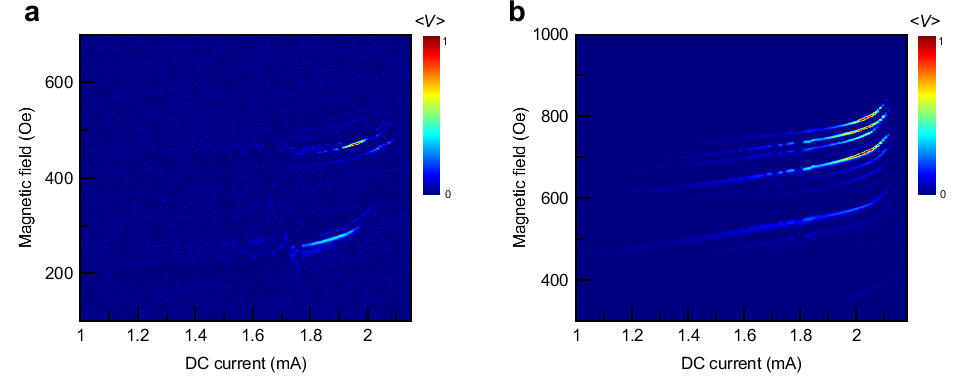}
\caption{Auto-oscillations of magnetization of the long nanowire. (a)~Microwave signal at 3.0\,GHz emitted from the nanowire in the easy axis. Normalized detected voltage $\langle V \rangle$ is proportional to the emitted microwave power density. (b)~Microwave emission in the hard axis. At low electric currents, a weak signal of thermally populated is visible. Above a current of approximately 1.8\,mA, microwave emission is strongly increased and the emission field shows a pronounced non-linear shift.}
\label{fig3}
\end{figure*}

The linewidth of the long nanowire in Fig.\,2c can be extrapolated to zero at about approximately 1.6\,mA, marking the critical current at which the intrinsic damping of the lowest spin wave mode is fully compensated by the spin Seebeck torque. Once the intrinsic damping is compensated, the magnetization transitions into auto-oscillations in the absence of external microwave drive\cite{slavin2009nonlinear}. The auto-oscillations are translated into electric signals via the effective magnetoresistance of the nanowire\cite{safranski2017spin}.

We supply electric DC current to the long nanowire and measure the microwave signal emitted from the device (into a microwave circuit\cite{safranski2017spin}, as described in Supporting Information). In Fig.\,3a, the signal detected at 3.0\,GHz is shown as a function of the magnetic field and DC current for the easy axis. We find that at low currents, the signal is nearly zero indicating only weak thermal excitation of the spin waves. Above the current of approximately 1.8\,mA, a sudden onset of the microwave signal is observed, which is characteristic of auto-oscillations\cite{slavin2009nonlinear}. By comparing the frequency-field relation of spin waves from auto-oscillations and from the ST-FMR experiment, we identify the lowest modes predominantly localized in the YIG layer to participate in auto-oscillations. With increasing current (around 2\,mA), the emission field shows a strong non-linear shift, which again is indicative of the auto-oscillatory regime\cite{slavin2009nonlinear,safranski2017spin,evelt2018spin}.

The microwave emission data, shown in Fig.\,3a for the easy axis and in Fig.\,3b for the hard axis, present similar critical currents for the onset of auto-oscillations. Inversion of magnetic field direction (Supporting Information) does not notably change the critical current, as well. The isotropy of the critical current further supports our conclusion -- in the long nanowire, isotropic\cite{safranski2017spin} thermal spin-torque due to spin Seebeck effect is strong and drives the auto-oscillations, while spin-orbit torques are significantly smaller. For the short nanowire, nearly free of the spin Seebeck effect, the critical current can be estimated to over 10\,mA, based on the ST-FMR linewidth in Fig.\,2c. Indeed, we observe no microwave emission from this device in the experimentally accessible current range.

The emergence of thermally driven magnetic auto-oscillations observed here can be understood as a manifestation of bosonic condensation of incoherent magnons into a coherent low-frequency magnon state. In order to reach the threshold of this instability, the thermal magnons need to be pumped to the chemical potential exceeding the natural frequency of the coherent low-frequency mode\cite{bender2014dynamic,demokritov2006bose,du2017control,demidov2017chemical}. Raising the chemical potential in one of the ferromagnet (here, YIG) by a heat flow from another one (Py), in two-magnet heterostructures, presents a conceptually novel scenario for magnetic auto-oscillations.

In Ref.\,\cite{bender2016thermally}, Bender et al. proposed a theoretical model for generating thermal spin currents between two ferromagnetic layers separated by a nonmagnetic metallic spacer. A temperature gradient across the layers injects a spin flow from the warmer ferromagnet into the spacer (raising its electronic spin accumulation), which is then transmitted downstream into the colder magnet. In our bi-layer system, the spacer is missing, so that the spin angular momentum is driven directly across the interface between the magnetic layers. Magnetic and possibly also magneto-elastic hybridization across the interface could result in an enhancement of the effective magnonic spin accumulation induced in the YIG layer, facilitating the onset of the condensation instability.

The underlying transfer of the spin angular momentum from the incoherent to the coherent spin precession can formally be also understood in terms of an anti-damping torque entering in the equation of motion for the latter, in proportion to the chemical potential of the incoherent magnon gas\cite{Flebus}. The chemical potential is zero in equilibrium but can be raised or lowered, depending on the orientation of the heat flux between the magnetic layers. We generically expect the hotter (colder) side to correspond to the decrease (increase) of the magnon chemical potential\cite{TwoFluid}.

Thermal spin-torque in heterogeneous systems, such as two-magnet systems investigated here, has so far remained unexplored experimentally and theoretically. We consider two possible microscopic mechanisms contributions to the torque: (i)~Spin waves are excitations of magnetic moments residing in the \textit{d}-electrons, which are coupled by \textit{s-d} exchange with conduction electrons in Py\cite{tserkovnyak2002spin}. The conduction electrons thus possess a higher temperature than spin waves which are delocalized over both, warmer and colder, layers. The conduction electrons serve as a source of energy and angular momentum and populate magnons in the nanowire beyond their thermal equilibrium, thus exerting a thermal spin-torque similar to the conventional spin Seebeck in YIG/Pt. In the two-magnet system, however, the spin Seebeck effect is different in that a temperature difference between the magnonic and electronic sub-systems within one heterogeneous magnetic system creates the anti-damping torque. (ii)~Since the spin waves at higher frequencies show a higher degree of delocalization, they are more susceptible to thermal excitations in the Py layer. The energy and angular momentum of higher-frequency magnons redistribute to the lower-frequency magnon states, creating a population beyond thermal equilibrium and exerting thermal spin-torque\cite{bender2014dynamic,demokritov2006bose,du2017control,demidov2017chemical}.

While our results demonstrate thermal spin-torque to be responsible for damping modification and auto-oscillations, they do not allow for quantitative comparison of contributions due to the (i)  magnon-electron and (ii) magnon-magnon mechanisms. However, the temperature gradient needed to reach auto-oscillations in our two-magnet nanowires and YIG/Pt nanowires of Ref.\,\cite{safranski2017spin} are similar. This allows for a conjecture that the magnon-magnon mechanism, which is only present in two-magnet systems, is unlikely more effective than the magnon-electron mechanism. The delocalized nature of spin waves offers an effective means for spin-charge conversion in applications based on magnetic insulators. We estimate the power spectral density of the emitted microwave signal from auto-oscillation to $\sim$1\,pW\,MHz$^{-1}$ which is three orders of magnitude higher than electrical signals obtained on YIG/Pt systems. 

In conclusion, we find that heterostructures based on a combination of metallic and insulating magnets bear a great promise for next-generation spintronic applications. Via interfacial spin coupling, magnetization dynamics in magnetic insulator translates into magnetization precession of the ferromagnetic metal. Inherently large magnetoresistance in ferromagnetic metal allows converting this magnetization dynamics into sizable electric signals. With a moderate interfacial spin coupling, the lower-energy spin wave modes are predominantly localized in one layer, while higher-energy modes show an increased level of hybridization and delocalization. Heat flow due to the temperature gradient across the layers leads to magnon condensation into the lower-energy modes, driving the magnetization into auto-oscillations. We demonstrate that a two-magnet system can be used to realize thermally driven spin-torque nano-oscillators. The results show a prospect for energy-efficient spintronic devices and advance our understanding of magnon thermodynamics in heterogeneous magnetic systems.  


\subsection{Associated Content}
\subsubsection{Supporting Information}
Fig.\,S1: Micromagnetic simulations of YIG/Py nanowire using MuMax3 package\cite{mumax}.\\
Fig.\,S2: Finite-element simulation (in COMSOL Multiphysics) of temperature profile under ohmic heating of the Py layer.\\
Fig.\,S3: Microwave emission from the long nanowire at negative fields.\\
Table\,S1: Overview of spin-orbit torques generated in ferromagnetic metals.

\subsection{Author Information}
\subsubsection{Corresponding Author}
*E-mail: igorb@ucr.edu

\subsubsection{ORCID}
Bassim Arkook: 0000-0002-4968-0954\\
Igor Barsukov: 0000-0003-4835-5620\\
Kilian Lenz: 0000-0001-5528-5080

\subsubsection{Author Contributions}
B.A. carried out magnetic characterization and performed data analysis. B.A and C.S. prepared the samples. R.R. and T.S. carried out micromagnetic simulations. H.C. and M.W. provided magnetic insulators. I.B. supervised the project. All authors discussed the data and co-wrote the manuscript.

\subsubsection{Notes}
The authors declare no competing financial interest.

\subsection{Acknowledgements}
This work was supported by the National Science Foundation under grant No.~ECCS-1810541. The material synthesis was supported as part of the ''Spins and Heat in Nanoscale Electronic Systems'' (SHINES), an Energy Frontier Research Center funded by the U.S. Department of Energy, Office of Science, Basic Energy Sciences (BES) under award No.~SC0012670. Work at CSU was supported by NSF through EFMA-1641989 and ECCS‐1915849. We thank NVIDIA Corporation for donating Titan Xp GPU used for some of the calculations. We thank Wangxiang Li for assistance in preparing the frontpage graphics.


\begin{mcitethebibliography}{47}
\providecommand*\natexlab[1]{#1}
\providecommand*\mciteSetBstSublistMode[1]{}
\providecommand*\mciteSetBstMaxWidthForm[2]{}
\providecommand*\mciteBstWouldAddEndPuncttrue
  {\def\EndOfBibitem{\unskip.}}
\providecommand*\mciteBstWouldAddEndPunctfalse
  {\let\EndOfBibitem\relax}
\providecommand*\mciteSetBstMidEndSepPunct[3]{}
\providecommand*\mciteSetBstSublistLabelBeginEnd[3]{}
\providecommand*\EndOfBibitem{}
\mciteSetBstSublistMode{f}
\mciteSetBstMaxWidthForm{subitem}{(\alph{mcitesubitemcount})}
\mciteSetBstSublistLabelBeginEnd
  {\mcitemaxwidthsubitemform\space}
  {\relax}
  {\relax}

\bibitem[Kiselev \latin{et~al.}(2003)Kiselev, Sankey, Krivorotov, Emley,
  Schoelkopf, Buhrman, and Ralph]{kiselev2003microwave}
Kiselev,~S.~I.; Sankey,~J.; Krivorotov,~I.; Emley,~N.; Schoelkopf,~R.;
  Buhrman,~R.; Ralph,~D. Microwave oscillations of a nanomagnet driven by a
  spin-polarized current. \emph{Nature} \textbf{2003}, \emph{425}, 380\relax
\mciteBstWouldAddEndPuncttrue
\mciteSetBstMidEndSepPunct{\mcitedefaultmidpunct}
{\mcitedefaultendpunct}{\mcitedefaultseppunct}\relax
\EndOfBibitem
\bibitem[Hellman \latin{et~al.}(2017)Hellman, Hoffmann, Tserkovnyak, Beach,
  Fullerton, Leighton, MacDonald, Ralph, Arena, D\"urr, Fischer, Grollier,
  Heremans, Jungwirth, Kimel, Koopmans, Krivorotov, May, Petford-Long,
  Rondinelli, Samarth, Schuller, Slavin, Stiles, Tchernyshyov, Thiaville, and
  Zink]{HellmanInterfaces}
Hellman,~F.; Hoffmann,~A.; Tserkovnyak,~Y.; Beach,~G. S.~D.; Fullerton,~E.~E.;
  Leighton,~C.; MacDonald,~A.~H.; Ralph,~D.~C.; Arena,~D.~A.; D\"urr,~H.~A.;
  Fischer,~P.; Grollier,~J.; Heremans,~J.~P.; Jungwirth,~T.; Kimel,~A.~V.;
  Koopmans,~B.; Krivorotov,~I.~N.; May,~S.~J.; Petford-Long,~A.~K.;
  Rondinelli,~J.~M.; Samarth,~N.; Schuller,~I.~K.; Slavin,~A.~N.;
  Stiles,~M.~D.; Tchernyshyov,~O.; Thiaville,~A.; Zink,~B.~L. Interface-induced
  phenomena in magnetism. \emph{Rev. Mod. Phys.} \textbf{2017}, \emph{89},
  025006\relax
\mciteBstWouldAddEndPuncttrue
\mciteSetBstMidEndSepPunct{\mcitedefaultmidpunct}
{\mcitedefaultendpunct}{\mcitedefaultseppunct}\relax
\EndOfBibitem
\bibitem[Avci \latin{et~al.}(2017)Avci, Quindeau, Pai, Mann, Caretta, Tang,
  Onbasli, Ross, and Beach]{avci2017current}
Avci,~C.~O.; Quindeau,~A.; Pai,~C.-F.; Mann,~M.; Caretta,~L.; Tang,~A.~S.;
  Onbasli,~M.~C.; Ross,~C.~A.; Beach,~G.~S. Current-induced switching in a
  magnetic insulator. \emph{Nat. Mater.} \textbf{2017}, \emph{16}, 309\relax
\mciteBstWouldAddEndPuncttrue
\mciteSetBstMidEndSepPunct{\mcitedefaultmidpunct}
{\mcitedefaultendpunct}{\mcitedefaultseppunct}\relax
\EndOfBibitem
\bibitem[Bozhko \latin{et~al.}(2016)Bozhko, Serga, Clausen, Vasyuchka,
  Heussner, Melkov, Pomyalov, L’vov, and Hillebrands]{bozhko2016supercurrent}
Bozhko,~D.~A.; Serga,~A.~A.; Clausen,~P.; Vasyuchka,~V.~I.; Heussner,~F.;
  Melkov,~G.~A.; Pomyalov,~A.; L’vov,~V.~S.; Hillebrands,~B. Supercurrent in
  a room-temperature Bose--Einstein magnon condensate. \emph{Nat. Phys.}
  \textbf{2016}, \emph{12}, 1057\relax
\mciteBstWouldAddEndPuncttrue
\mciteSetBstMidEndSepPunct{\mcitedefaultmidpunct}
{\mcitedefaultendpunct}{\mcitedefaultseppunct}\relax
\EndOfBibitem
\bibitem[Yang and Hammel(2018)Yang, and Hammel]{yang2018fmr}
Yang,~F.; Hammel,~P.~C. FMR-driven spin pumping in Y3Fe5O12-based structures.
  \emph{J. Phys. D: Appl. Phys.} \textbf{2018}, \emph{51}, 253001\relax
\mciteBstWouldAddEndPuncttrue
\mciteSetBstMidEndSepPunct{\mcitedefaultmidpunct}
{\mcitedefaultendpunct}{\mcitedefaultseppunct}\relax
\EndOfBibitem
\bibitem[Wunderlich(2017)]{wunderlich2017spintronics}
Wunderlich,~J. Spintronics: Current-switched magnetic insulator. \emph{Nat.
  Mater.} \textbf{2017}, \emph{16}, 284\relax
\mciteBstWouldAddEndPuncttrue
\mciteSetBstMidEndSepPunct{\mcitedefaultmidpunct}
{\mcitedefaultendpunct}{\mcitedefaultseppunct}\relax
\EndOfBibitem
\bibitem[Collet \latin{et~al.}(2016)Collet, De~Milly, Kelly, Naletov, Bernard,
  Bortolotti, Youssef, Demidov, Demokritov, Prieto, Munoz, Cros, Anane, and
  Klein]{collet2016generation}
Collet,~M.; De~Milly,~X.; Kelly,~O.~d.; Naletov,~V.~V.; Bernard,~R.;
  Bortolotti,~P.; Youssef,~J.~B.; Demidov,~V.; Demokritov,~S.; Prieto,~J.;
  Munoz,~M.; Cros,~V.; Anane,~G.,~A. adn de~Loubens; Klein,~O. Generation of
  coherent spin-wave modes in yttrium iron garnet microdiscs by spin--orbit
  torque. \emph{Nat. Commun.} \textbf{2016}, \emph{7}, 10377\relax
\mciteBstWouldAddEndPuncttrue
\mciteSetBstMidEndSepPunct{\mcitedefaultmidpunct}
{\mcitedefaultendpunct}{\mcitedefaultseppunct}\relax
\EndOfBibitem
\bibitem[Safranski \latin{et~al.}(2017)Safranski, Barsukov, Lee, Schneider,
  Jara, Smith, Chang, Lenz, Lindner, Tserkovnyak, Wu, and
  Krivorotov]{safranski2017spin}
Safranski,~C.; Barsukov,~I.; Lee,~H.~K.; Schneider,~T.; Jara,~A.; Smith,~A.;
  Chang,~H.; Lenz,~K.; Lindner,~J.; Tserkovnyak,~Y.; Wu,~M.; Krivorotov,~I.~N.
  Spin caloritronic nano-oscillator. \emph{Nat. Commun.} \textbf{2017},
  \emph{8}, 117\relax
\mciteBstWouldAddEndPuncttrue
\mciteSetBstMidEndSepPunct{\mcitedefaultmidpunct}
{\mcitedefaultendpunct}{\mcitedefaultseppunct}\relax
\EndOfBibitem
\bibitem[Lee \latin{et~al.}(2016)Lee, Barsukov, Swartz, Kim, Yang, Hwang, and
  Krivorotov]{lee2016magnetic}
Lee,~H.~K.; Barsukov,~I.; Swartz,~A.; Kim,~B.; Yang,~L.; Hwang,~H.;
  Krivorotov,~I. Magnetic anisotropy, damping, and interfacial spin transport
  in Pt/LSMO bilayers. \emph{AIP Advances} \textbf{2016}, \emph{6},
  055212\relax
\mciteBstWouldAddEndPuncttrue
\mciteSetBstMidEndSepPunct{\mcitedefaultmidpunct}
{\mcitedefaultendpunct}{\mcitedefaultseppunct}\relax
\EndOfBibitem
\bibitem[Nakayama \latin{et~al.}(2013)Nakayama, Althammer, Chen, Uchida,
  Kajiwara, Kikuchi, Ohtani, Gepr\"ags, Opel, Takahashi, Gross, Bauer,
  Goennenwein, and Saitoh]{nakayama2013spin}
Nakayama,~H.; Althammer,~M.; Chen,~Y.-T.; Uchida,~K.; Kajiwara,~Y.;
  Kikuchi,~D.; Ohtani,~T.; Gepr\"ags,~S.; Opel,~M.; Takahashi,~S.; Gross,~R.;
  Bauer,~G. E.~W.; Goennenwein,~S. T.~B.; Saitoh,~E. Spin Hall
  Magnetoresistance Induced by a Nonequilibrium Proximity Effect. \emph{Phys.
  Rev. Lett.} \textbf{2013}, \emph{110}, 206601\relax
\mciteBstWouldAddEndPuncttrue
\mciteSetBstMidEndSepPunct{\mcitedefaultmidpunct}
{\mcitedefaultendpunct}{\mcitedefaultseppunct}\relax
\EndOfBibitem
\bibitem[Sinova \latin{et~al.}(2015)Sinova, Valenzuela, Wunderlich, Back, and
  Jungwirth]{SinovaSHE}
Sinova,~J.; Valenzuela,~S.~O.; Wunderlich,~J.; Back,~C.~H.; Jungwirth,~T. Spin
  Hall effects. \emph{Rev. Mod. Phys.} \textbf{2015}, \emph{87},
  1213--1260\relax
\mciteBstWouldAddEndPuncttrue
\mciteSetBstMidEndSepPunct{\mcitedefaultmidpunct}
{\mcitedefaultendpunct}{\mcitedefaultseppunct}\relax
\EndOfBibitem
\bibitem[Tulapurkar \latin{et~al.}(2005)Tulapurkar, Suzuki, Fukushima, Kubota,
  Maehara, Tsunekawa, Djayaprawira, Watanabe, and Yuasa]{tulapurkar2005spin}
Tulapurkar,~A.; Suzuki,~Y.; Fukushima,~A.; Kubota,~H.; Maehara,~H.;
  Tsunekawa,~K.; Djayaprawira,~D.; Watanabe,~N.; Yuasa,~S. Spin-torque diode
  effect in magnetic tunnel junctions. \emph{Nature} \textbf{2005}, \emph{438},
  339\relax
\mciteBstWouldAddEndPuncttrue
\mciteSetBstMidEndSepPunct{\mcitedefaultmidpunct}
{\mcitedefaultendpunct}{\mcitedefaultseppunct}\relax
\EndOfBibitem
\bibitem[Baumgaertl \latin{et~al.}(2016)Baumgaertl, Heimbach, Maendl, Rueffer,
  Fontcuberta~i Morral, and Grundler]{baumgaertl2016magnetization}
Baumgaertl,~K.; Heimbach,~F.; Maendl,~S.; Rueffer,~D.; Fontcuberta~i
  Morral,~A.; Grundler,~D. Magnetization reversal in individual Py and CoFeB
  nanotubes locally probed via anisotropic magnetoresistance and anomalous
  Nernst effect. \emph{Appl. Phys. Lett.} \textbf{2016}, \emph{108},
  132408\relax
\mciteBstWouldAddEndPuncttrue
\mciteSetBstMidEndSepPunct{\mcitedefaultmidpunct}
{\mcitedefaultendpunct}{\mcitedefaultseppunct}\relax
\EndOfBibitem
\bibitem[Taniguchi \latin{et~al.}(2015)Taniguchi, Grollier, and
  Stiles]{taniguchi2015spin}
Taniguchi,~T.; Grollier,~J.; Stiles,~M.~D. Spin-Transfer Torques Generated by
  the Anomalous Hall Effect and Anisotropic Magnetoresistance. \emph{Phys. Rev.
  Appl.} \textbf{2015}, \emph{3}, 044001\relax
\mciteBstWouldAddEndPuncttrue
\mciteSetBstMidEndSepPunct{\mcitedefaultmidpunct}
{\mcitedefaultendpunct}{\mcitedefaultseppunct}\relax
\EndOfBibitem
\bibitem[Davidson \latin{et~al.}(2019)Davidson, Amin, Aljuaid, Haney, and
  Fan]{davidson2019perspectives}
Davidson,~A.; Amin,~V.~P.; Aljuaid,~W.~S.; Haney,~P.~M.; Fan,~X. Perspectives
  of Electrically generated spin currents in ferromagnetic materials.
  \emph{arXiv:1906.11772} \textbf{2019}, \relax
\mciteBstWouldAddEndPunctfalse
\mciteSetBstMidEndSepPunct{\mcitedefaultmidpunct}
{}{\mcitedefaultseppunct}\relax
\EndOfBibitem
\bibitem[Amin \latin{et~al.}(2018)Amin, Zemen, and Stiles]{AminInterface}
Amin,~V.~P.; Zemen,~J.; Stiles,~M.~D. Interface-Generated Spin Currents.
  \emph{Phys. Rev. Lett.} \textbf{2018}, \emph{121}, 136805\relax
\mciteBstWouldAddEndPuncttrue
\mciteSetBstMidEndSepPunct{\mcitedefaultmidpunct}
{\mcitedefaultendpunct}{\mcitedefaultseppunct}\relax
\EndOfBibitem
\bibitem[Gibbons \latin{et~al.}(2018)Gibbons, MacNeill, Buhrman, and
  Ralph]{RalphAHE}
Gibbons,~J.~D.; MacNeill,~D.; Buhrman,~R.~A.; Ralph,~D.~C. Reorientable Spin
  Direction for Spin Current Produced by the Anomalous Hall Effect. \emph{Phys.
  Rev. Appl.} \textbf{2018}, \emph{9}, 064033\relax
\mciteBstWouldAddEndPuncttrue
\mciteSetBstMidEndSepPunct{\mcitedefaultmidpunct}
{\mcitedefaultendpunct}{\mcitedefaultseppunct}\relax
\EndOfBibitem
\bibitem[Das \latin{et~al.}(2018)Das, Liu, van Wees, and Vera-Marun]{DasAHE}
Das,~K.~S.; Liu,~J.; van Wees,~B.~J.; Vera-Marun,~I.~J. Efficient Injection and
  Detection of Out-of-Plane Spins via the Anomalous Spin Hall Effect in
  Permalloy Nanowires. \emph{Nano Lett.} \textbf{2018}, \emph{18},
  5633--5639\relax
\mciteBstWouldAddEndPuncttrue
\mciteSetBstMidEndSepPunct{\mcitedefaultmidpunct}
{\mcitedefaultendpunct}{\mcitedefaultseppunct}\relax
\EndOfBibitem
\bibitem[Bose \latin{et~al.}(2018)Bose, Lam, Bhuktare, Dutta, Singh, Jibiki,
  Goto, Miwa, and Tulapurkar]{TulapurkarAHE}
Bose,~A.; Lam,~D.~D.; Bhuktare,~S.; Dutta,~S.; Singh,~H.; Jibiki,~Y.; Goto,~M.;
  Miwa,~S.; Tulapurkar,~A.~A. Observation of Anomalous Spin Torque Generated by
  a Ferromagnet. \emph{Phys. Rev. Appl.} \textbf{2018}, \emph{9}, 064026\relax
\mciteBstWouldAddEndPuncttrue
\mciteSetBstMidEndSepPunct{\mcitedefaultmidpunct}
{\mcitedefaultendpunct}{\mcitedefaultseppunct}\relax
\EndOfBibitem
\bibitem[Safranski \latin{et~al.}(2019)Safranski, Montoya, and
  Krivorotov]{safranski2019spin}
Safranski,~C.; Montoya,~E.~A.; Krivorotov,~I.~N. Spin--orbit torque driven by a
  planar hall current. \emph{Nat. Nanotech.} \textbf{2019}, \emph{14}, 27\relax
\mciteBstWouldAddEndPuncttrue
\mciteSetBstMidEndSepPunct{\mcitedefaultmidpunct}
{\mcitedefaultendpunct}{\mcitedefaultseppunct}\relax
\EndOfBibitem
\bibitem[Humphries \latin{et~al.}(2017)Humphries, Wang, Edwards, Allen, Shaw,
  Nembach, Xiao, Silva, and Fan]{HumphriesRotSym}
Humphries,~A.~M.; Wang,~T.; Edwards,~E.~R.; Allen,~S.~R.; Shaw,~J.~M.;
  Nembach,~H.~T.; Xiao,~J.~Q.; Silva,~T.~J.; Fan,~X. Observation of spin-orbit
  effects with spin rotation symmetry. \emph{Nat. Commun.} \textbf{2017},
  \emph{8}, 911\relax
\mciteBstWouldAddEndPuncttrue
\mciteSetBstMidEndSepPunct{\mcitedefaultmidpunct}
{\mcitedefaultendpunct}{\mcitedefaultseppunct}\relax
\EndOfBibitem
\bibitem[Haidar \latin{et~al.}(2019)Haidar, Awad, Dvornik, Khymyn, Houshang,
  and {\AA}kerman]{haidar2019single}
Haidar,~M.; Awad,~A.~A.; Dvornik,~M.; Khymyn,~R.; Houshang,~A.; {\AA}kerman,~J.
  A single layer spin-orbit torque nano-oscillator. \emph{Nat. Commun.}
  \textbf{2019}, \emph{10}, 2362\relax
\mciteBstWouldAddEndPuncttrue
\mciteSetBstMidEndSepPunct{\mcitedefaultmidpunct}
{\mcitedefaultendpunct}{\mcitedefaultseppunct}\relax
\EndOfBibitem
\bibitem[Demokritov \latin{et~al.}(2006)Demokritov, Demidov, Dzyapko, Melkov,
  Serga, Hillebrands, and Slavin]{demokritov2006bose}
Demokritov,~S.; Demidov,~V.; Dzyapko,~O.; Melkov,~G.; Serga,~A.;
  Hillebrands,~B.; Slavin,~A. Bose--Einstein condensation of quasi-equilibrium
  magnons at room temperature under pumping. \emph{Nature} \textbf{2006},
  \emph{443}, 430\relax
\mciteBstWouldAddEndPuncttrue
\mciteSetBstMidEndSepPunct{\mcitedefaultmidpunct}
{\mcitedefaultendpunct}{\mcitedefaultseppunct}\relax
\EndOfBibitem
\bibitem[Liu \latin{et~al.}(2014)Liu, Chang, Vlaminck, Sun, Kabatek, Hoffmann,
  Deng, and Wu]{liu2014ferromagnetic}
Liu,~T.; Chang,~H.; Vlaminck,~V.; Sun,~Y.; Kabatek,~M.; Hoffmann,~A.; Deng,~L.;
  Wu,~M. Ferromagnetic resonance of sputtered yttrium iron garnet nanometer
  films. \emph{J. Appl. Phys.} \textbf{2014}, \emph{115}, 17A501\relax
\mciteBstWouldAddEndPuncttrue
\mciteSetBstMidEndSepPunct{\mcitedefaultmidpunct}
{\mcitedefaultendpunct}{\mcitedefaultseppunct}\relax
\EndOfBibitem
\bibitem[Duan \latin{et~al.}(2014)Duan, Smith, Yang, Youngblood, Lindner,
  Demidov, Demokritov, and Krivorotov]{duan2014nanowire}
Duan,~Z.; Smith,~A.; Yang,~L.; Youngblood,~B.; Lindner,~J.; Demidov,~V.~E.;
  Demokritov,~S.~O.; Krivorotov,~I.~N. Nanowire spin torque oscillator driven
  by spin orbit torques. \emph{Nat. Commun.} \textbf{2014}, \emph{5},
  5616\relax
\mciteBstWouldAddEndPuncttrue
\mciteSetBstMidEndSepPunct{\mcitedefaultmidpunct}
{\mcitedefaultendpunct}{\mcitedefaultseppunct}\relax
\EndOfBibitem
\bibitem[Sluka \latin{et~al.}(2019)Sluka, Schneider, Gallardo, K{\'a}kay,
  Weigand, Warnatz, Mattheis, Rold{\'a}n-Molina, Landeros, Tiberkevich, Slavin,
  Schütz, Erbe, Deac, Jürgen~Lindner, Raabe, Fassbender, and
  Wintz]{sluka2019emission}
Sluka,~V.; Schneider,~T.; Gallardo,~R.~A.; K{\'a}kay,~A.; Weigand,~M.;
  Warnatz,~T.; Mattheis,~R.; Rold{\'a}n-Molina,~A.; Landeros,~P.;
  Tiberkevich,~V.; Slavin,~A.; Schütz,~G.; Erbe,~A.; Deac,~A.;
  Jürgen~Lindner,~J.; Raabe,~J.; Fassbender,~J.; Wintz,~S. Emission and
  propagation of 1D and 2D spin waves with nanoscale wavelengths in anisotropic
  spin textures. \emph{Nat. Nanotech.} \textbf{2019}, \emph{14}, 328\relax
\mciteBstWouldAddEndPuncttrue
\mciteSetBstMidEndSepPunct{\mcitedefaultmidpunct}
{\mcitedefaultendpunct}{\mcitedefaultseppunct}\relax
\EndOfBibitem
\bibitem[An \latin{et~al.}(2019)An, Bhat, Mruczkiewicz, Dubs, and
  Grundler]{PhysRevApplied.11.034065}
An,~K.; Bhat,~V.; Mruczkiewicz,~M.; Dubs,~C.; Grundler,~D. Optimization of
  Spin-Wave Propagation with Enhanced Group Velocities by Exchange-Coupled
  Ferrimagnet-Ferromagnet Bilayers. \emph{Phys. Rev. Appl.} \textbf{2019},
  \emph{11}, 034065\relax
\mciteBstWouldAddEndPuncttrue
\mciteSetBstMidEndSepPunct{\mcitedefaultmidpunct}
{\mcitedefaultendpunct}{\mcitedefaultseppunct}\relax
\EndOfBibitem
\bibitem[Liu \latin{et~al.}(2018)Liu, Chen, Liu, Heimbach, Yu, Xiao, Hu, Liu,
  Chang, Stueckler, Tu, Zhang, Zhang, Gao, Liao, Yu, Xia, Zhao, and
  Wu]{liu2018long}
Liu,~C.; Chen,~J.; Liu,~T.; Heimbach,~F.; Yu,~H.; Xiao,~Y.; Hu,~J.; Liu,~M.;
  Chang,~H.; Stueckler,~T.; Tu,~S.; Zhang,~Y.; Zhang,~Y.; Gao,~P.; Liao,~Z.;
  Yu,~D.; Xia,~N.,~Ke~Lei; Zhao,~W.; Wu,~M. Long-distance propagation of
  short-wavelength spin waves. \emph{Nat. Commun.} \textbf{2018}, \emph{9},
  738\relax
\mciteBstWouldAddEndPuncttrue
\mciteSetBstMidEndSepPunct{\mcitedefaultmidpunct}
{\mcitedefaultendpunct}{\mcitedefaultseppunct}\relax
\EndOfBibitem
\bibitem[Cochran \latin{et~al.}(1986)Cochran, Heinrich, and
  Arrott]{PhysRevB.34.7788}
Cochran,~J.~F.; Heinrich,~B.; Arrott,~A.~S. Ferromagnetic resonance in a system
  composed of a ferromagnetic substrate and an exchange-coupled thin
  ferromagnetic overlayer. \emph{Phys. Rev. B} \textbf{1986}, \emph{34},
  7788--7801\relax
\mciteBstWouldAddEndPuncttrue
\mciteSetBstMidEndSepPunct{\mcitedefaultmidpunct}
{\mcitedefaultendpunct}{\mcitedefaultseppunct}\relax
\EndOfBibitem
\bibitem[Nogués and Schuller(1999)Nogués, and Schuller]{NOGUES1999203}
Nogués,~J.; Schuller,~I.~K. Exchange bias. \emph{J. Magn. Magn. Mater.}
  \textbf{1999}, \emph{192}, 203 -- 232\relax
\mciteBstWouldAddEndPuncttrue
\mciteSetBstMidEndSepPunct{\mcitedefaultmidpunct}
{\mcitedefaultendpunct}{\mcitedefaultseppunct}\relax
\EndOfBibitem
\bibitem[Klingler \latin{et~al.}(2018)Klingler, Amin, Gepr\"ags, Ganzhorn,
  Maier-Flaig, Althammer, Huebl, Gross, McMichael, Stiles, Goennenwein, and
  Weiler]{PhysRevLett.120.127201}
Klingler,~S.; Amin,~V.; Gepr\"ags,~S.; Ganzhorn,~K.; Maier-Flaig,~H.;
  Althammer,~M.; Huebl,~H.; Gross,~R.; McMichael,~R.~D.; Stiles,~M.~D.;
  Goennenwein,~S. T.~B.; Weiler,~M. Spin-Torque Excitation of Perpendicular
  Standing Spin Waves in Coupled $\mathrm{YIG}/\mathrm{Co}$ Heterostructures.
  \emph{Phys. Rev. Lett.} \textbf{2018}, \emph{120}, 127201\relax
\mciteBstWouldAddEndPuncttrue
\mciteSetBstMidEndSepPunct{\mcitedefaultmidpunct}
{\mcitedefaultendpunct}{\mcitedefaultseppunct}\relax
\EndOfBibitem
\bibitem[Chen \latin{et~al.}(2018)Chen, Liu, Liu, Xiao, Xia, Bauer, Wu, and
  Yu]{PhysRevLett.120.217202}
Chen,~J.; Liu,~C.; Liu,~T.; Xiao,~Y.; Xia,~K.; Bauer,~G. E.~W.; Wu,~M.; Yu,~H.
  Strong Interlayer Magnon-Magnon Coupling in Magnetic Metal-Insulator Hybrid
  Nanostructures. \emph{Phys. Rev. Lett.} \textbf{2018}, \emph{120},
  217202\relax
\mciteBstWouldAddEndPuncttrue
\mciteSetBstMidEndSepPunct{\mcitedefaultmidpunct}
{\mcitedefaultendpunct}{\mcitedefaultseppunct}\relax
\EndOfBibitem
\bibitem[Gon{\c{c}}alves \latin{et~al.}(2013)Gon{\c{c}}alves, Barsukov, Chen,
  Yang, Katine, and Krivorotov]{gonccalves2013spin}
Gon{\c{c}}alves,~A.; Barsukov,~I.; Chen,~Y.-J.; Yang,~L.; Katine,~J.;
  Krivorotov,~I. Spin torque ferromagnetic resonance with magnetic field
  modulation. \emph{Appl. Phys. Lett.} \textbf{2013}, \emph{103}, 172406\relax
\mciteBstWouldAddEndPuncttrue
\mciteSetBstMidEndSepPunct{\mcitedefaultmidpunct}
{\mcitedefaultendpunct}{\mcitedefaultseppunct}\relax
\EndOfBibitem
\bibitem[Duan \latin{et~al.}(2015)Duan, Krivorotov, Arias, Reckers, Stienen,
  and Lindner]{PhysRevB.92.104424}
Duan,~Z.; Krivorotov,~I.~N.; Arias,~R.~E.; Reckers,~N.; Stienen,~S.;
  Lindner,~J. Spin wave eigenmodes in transversely magnetized thin film
  ferromagnetic wires. \emph{Phys. Rev. B} \textbf{2015}, \emph{92},
  104424\relax
\mciteBstWouldAddEndPuncttrue
\mciteSetBstMidEndSepPunct{\mcitedefaultmidpunct}
{\mcitedefaultendpunct}{\mcitedefaultseppunct}\relax
\EndOfBibitem
\bibitem[Ohnuma \latin{et~al.}(2015)Ohnuma, Adachi, Saitoh, and
  Maekawa]{ohnuma2015magnon}
Ohnuma,~Y.; Adachi,~H.; Saitoh,~E.; Maekawa,~S. Magnon instability driven by
  heat current in magnetic bilayers. \emph{Phys. Rev. B} \textbf{2015},
  \emph{92}, 224404\relax
\mciteBstWouldAddEndPuncttrue
\mciteSetBstMidEndSepPunct{\mcitedefaultmidpunct}
{\mcitedefaultendpunct}{\mcitedefaultseppunct}\relax
\EndOfBibitem
\bibitem[Bender \latin{et~al.}(2014)Bender, Duine, Brataas, and
  Tserkovnyak]{bender2014dynamic}
Bender,~S.~A.; Duine,~R.~A.; Brataas,~A.; Tserkovnyak,~Y. Dynamic phase diagram
  of dc-pumped magnon condensates. \emph{Phys. Rev. B} \textbf{2014},
  \emph{90}, 094409\relax
\mciteBstWouldAddEndPuncttrue
\mciteSetBstMidEndSepPunct{\mcitedefaultmidpunct}
{\mcitedefaultendpunct}{\mcitedefaultseppunct}\relax
\EndOfBibitem
\bibitem[Lu \latin{et~al.}(2012)Lu, Sun, Jantz, and Wu]{WuSpinSeebeckPRL}
Lu,~L.; Sun,~Y.; Jantz,~M.; Wu,~M. Control of Ferromagnetic Relaxation in
  Magnetic Thin Films through Thermally Induced Interfacial Spin Transfer.
  \emph{Phys. Rev. Lett.} \textbf{2012}, \emph{108}, 257202\relax
\mciteBstWouldAddEndPuncttrue
\mciteSetBstMidEndSepPunct{\mcitedefaultmidpunct}
{\mcitedefaultendpunct}{\mcitedefaultseppunct}\relax
\EndOfBibitem
\bibitem[Slavin and Tiberkevich(2009)Slavin, and
  Tiberkevich]{slavin2009nonlinear}
Slavin,~A.; Tiberkevich,~V. Nonlinear auto-oscillator theory of microwave
  generation by spin-polarized current. \emph{IEEE Trans. Magn.} \textbf{2009},
  \emph{45}, 1875--1918\relax
\mciteBstWouldAddEndPuncttrue
\mciteSetBstMidEndSepPunct{\mcitedefaultmidpunct}
{\mcitedefaultendpunct}{\mcitedefaultseppunct}\relax
\EndOfBibitem
\bibitem[Evelt \latin{et~al.}(2018)Evelt, Safranski, Aldosary, Demidov,
  Barsukov, Nosov, Rinkevich, Sobotkiewich, Li, Shi, Krivorotov, and
  Demokritov]{evelt2018spin}
Evelt,~M.; Safranski,~C.; Aldosary,~M.; Demidov,~V.; Barsukov,~I.; Nosov,~A.;
  Rinkevich,~A.; Sobotkiewich,~K.; Li,~X.; Shi,~J.; Krivorotov,~I.~N.;
  Demokritov,~S.~O. Spin Hall-induced auto-oscillations in ultrathin YIG grown
  on Pt. \emph{Sci. Rep.} \textbf{2018}, \emph{8}, 1269\relax
\mciteBstWouldAddEndPuncttrue
\mciteSetBstMidEndSepPunct{\mcitedefaultmidpunct}
{\mcitedefaultendpunct}{\mcitedefaultseppunct}\relax
\EndOfBibitem
\bibitem[Du \latin{et~al.}(2017)Du, Van~der Sar, Zhou, Upadhyaya, Casola,
  Zhang, Onbasli, Ross, Walsworth, Tserkovnyak, and Yacoby]{du2017control}
Du,~C.; Van~der Sar,~T.; Zhou,~T.~X.; Upadhyaya,~P.; Casola,~F.; Zhang,~H.;
  Onbasli,~M.~C.; Ross,~C.~A.; Walsworth,~R.~L.; Tserkovnyak,~Y.; Yacoby,~A.
  Control and local measurement of the spin chemical potential in a magnetic
  insulator. \emph{Science} \textbf{2017}, \emph{357}, 195--198\relax
\mciteBstWouldAddEndPuncttrue
\mciteSetBstMidEndSepPunct{\mcitedefaultmidpunct}
{\mcitedefaultendpunct}{\mcitedefaultseppunct}\relax
\EndOfBibitem
\bibitem[Demidov \latin{et~al.}(2017)Demidov, Urazhdin, Divinskiy, Bessonov,
  Rinkevich, Ustinov, and Demokritov]{demidov2017chemical}
Demidov,~V.; Urazhdin,~S.; Divinskiy,~B.; Bessonov,~V.; Rinkevich,~A.;
  Ustinov,~V.; Demokritov,~S. Chemical potential of quasi-equilibrium magnon
  gas driven by pure spin current. \emph{Nat. Commun.} \textbf{2017}, \emph{8},
  1579\relax
\mciteBstWouldAddEndPuncttrue
\mciteSetBstMidEndSepPunct{\mcitedefaultmidpunct}
{\mcitedefaultendpunct}{\mcitedefaultseppunct}\relax
\EndOfBibitem
\bibitem[Bender and Tserkovnyak(2016)Bender, and
  Tserkovnyak]{bender2016thermally}
Bender,~S.~A.; Tserkovnyak,~Y. Thermally driven spin torques in layered
  magnetic insulators. \emph{Phys. Rev. B} \textbf{2016}, \emph{93},
  064418\relax
\mciteBstWouldAddEndPuncttrue
\mciteSetBstMidEndSepPunct{\mcitedefaultmidpunct}
{\mcitedefaultendpunct}{\mcitedefaultseppunct}\relax
\EndOfBibitem
\bibitem[Flebus \latin{et~al.}(2016)Flebus, Upadhyaya, Duine, and
  Tserkovnyak]{Flebus}
Flebus,~B.; Upadhyaya,~P.; Duine,~R.~A.; Tserkovnyak,~Y. Local thermomagnonic
  torques in two-fluid spin dynamics. \emph{Phys. Rev. B} \textbf{2016},
  \emph{94}, 214428\relax
\mciteBstWouldAddEndPuncttrue
\mciteSetBstMidEndSepPunct{\mcitedefaultmidpunct}
{\mcitedefaultendpunct}{\mcitedefaultseppunct}\relax
\EndOfBibitem
\bibitem[Flebus \latin{et~al.}(2016)Flebus, Bender, Tserkovnyak, and
  Duine]{TwoFluid}
Flebus,~B.; Bender,~S.~A.; Tserkovnyak,~Y.; Duine,~R.~A. Two-Fluid Theory for
  Spin Superfluidity in Magnetic Insulators. \emph{Phys. Rev. Lett.}
  \textbf{2016}, \emph{116}, 117201\relax
\mciteBstWouldAddEndPuncttrue
\mciteSetBstMidEndSepPunct{\mcitedefaultmidpunct}
{\mcitedefaultendpunct}{\mcitedefaultseppunct}\relax
\EndOfBibitem
\bibitem[Tserkovnyak \latin{et~al.}(2002)Tserkovnyak, Brataas, and
  Bauer]{tserkovnyak2002spin}
Tserkovnyak,~Y.; Brataas,~A.; Bauer,~G. E.~W. Spin pumping and magnetization
  dynamics in metallic multilayers. \emph{Phys. Rev. B} \textbf{2002},
  \emph{66}, 224403\relax
\mciteBstWouldAddEndPuncttrue
\mciteSetBstMidEndSepPunct{\mcitedefaultmidpunct}
{\mcitedefaultendpunct}{\mcitedefaultseppunct}\relax
\EndOfBibitem
\bibitem[Vansteenkiste \latin{et~al.}(2014)Vansteenkiste, Leliaert, Dvornik,
  Helsen, Garcia-Sanchez, and Van~Waeyenberge]{mumax}
Vansteenkiste,~A.; Leliaert,~J.; Dvornik,~M.; Helsen,~M.; Garcia-Sanchez,~F.;
  Van~Waeyenberge,~B. The design and verification of MuMax3. \emph{AIP Adv.}
  \textbf{2014}, \emph{4}, 107133\relax
\mciteBstWouldAddEndPuncttrue
\mciteSetBstMidEndSepPunct{\mcitedefaultmidpunct}
{\mcitedefaultendpunct}{\mcitedefaultseppunct}\relax
\EndOfBibitem
\end{mcitethebibliography}

\newpage
\includepdf[pages={1}]{empty}
~
\newpage
\includepdf[pages={1}]{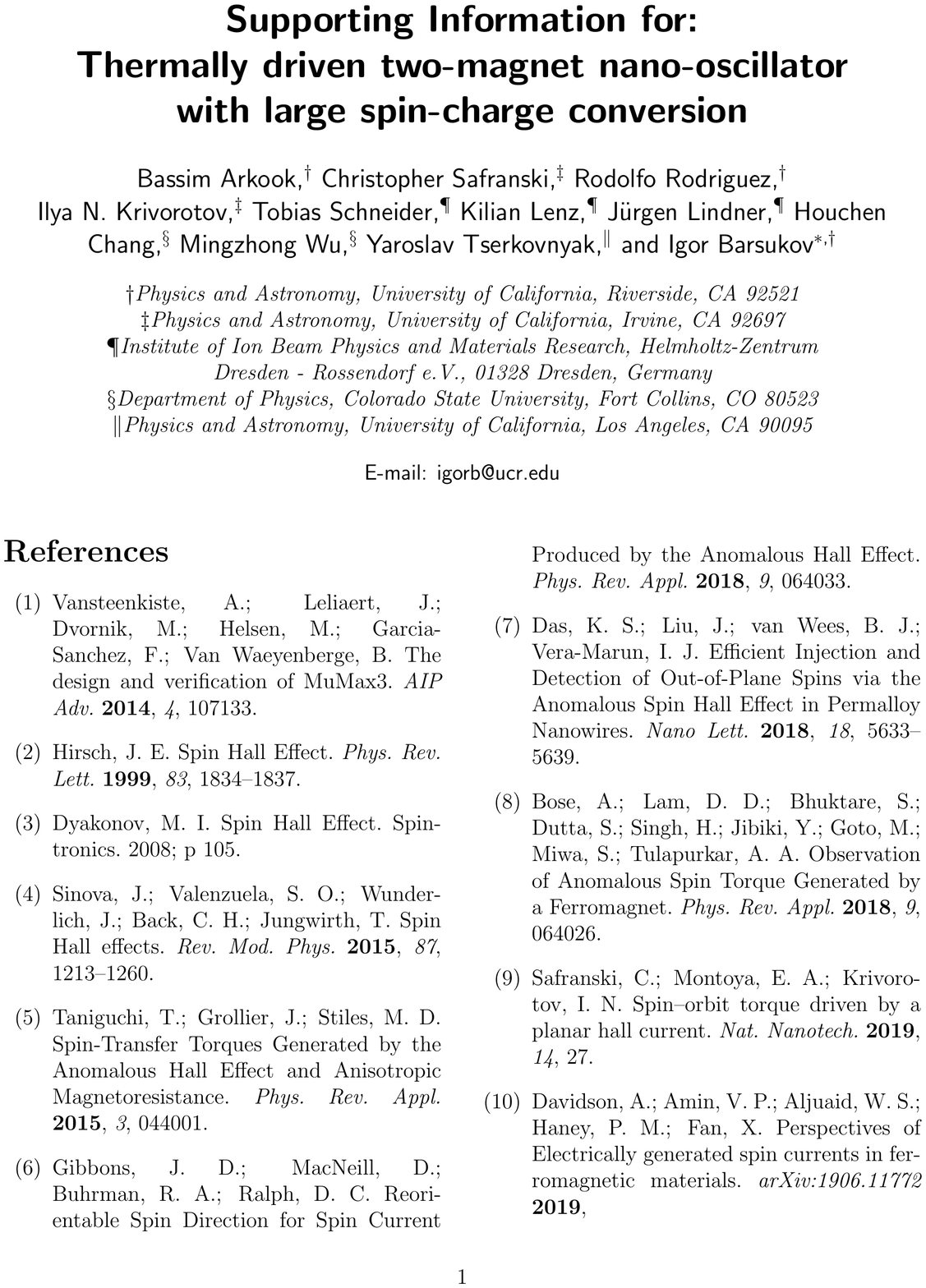}
~
\newpage
\includepdf[pages={2}]{supplemental}
~
\newpage
\includepdf[pages={3}]{supplemental}
~
\newpage
\includepdf[pages={4}]{supplemental}
~
\newpage
\includepdf[pages={5}]{supplemental}

\end{document}